# On the problem of a physical "theory of everything"

## Jerzy Klemens Kowalczyński

Retired from the Institute of Physics, Polish Academy of Sciences

Correspondence address: J. K. Kowalczyński, ON3, Institute of Physics, Polish Academy of Sciences, Al. Lotników 32/46, 02-668 Warsaw, Poland

E-mail: jkowal@ifpan.edu.pl

**Abstract:** No quantitative theory describing all physical phenomena can be made if any arbitrary standard spacetime structure is assumed. This statement is a consequence of transforming the Peano arithmetic axioms into sentences with a physical content.

Some physicists imagine and some others even believe that a theory describing all physical phenomena – briefly called a "theory of everything" – may exist. An amusing derivative of such an attitude is the use of this term by some relativists who try to produce a quantum theory of gravity. (This trend, recently very modish in general relativity, seems to be quite unfortunate since quantum physics is predisposed to describe the microworld where the gravitational interactions are negligible.)

On a more serious note, let us note that today the term "physical theories" or more precisely: "quantitative physical theories" – is understood to mean descriptions (models) of physical phenomena in the languages of mathematical theories. Almost all the present-day physical theories are based upon essentially incomplete mathematical theories. The fact that descriptions of even simple physical phenomena need such strong mathematical tools should immediately lead to the conclusion that any "theory of everything" cannot be made, since we shall always encounter independent sentences with a physical content.

Defenders of the "theory of everything" idea might argue that we can try to describe physical phenomena in terms of complete mathematical theories. This, however, would mean creation of the whole theoretical physics anew by using very poor mathematical tools. They might also argue that the following situation is not excluded: all the physical phenomena are described by a set of provable sentences, though some (or all) of these sentences are expressed in terms of essentially incomplete mathematical theories. It is shown below that these situations cannot be realized if any arbitrary standard spacetime structure is assumed.

The following sentences

- A1. $0 \in N$
- A2. $\forall_{n \in N} \exists_{m \in N} [m = S(n)]$
- A3. $\forall_{n \in N} [0 \neq S(n)]$
- A4. $\forall_{n \in N} \forall_{m \in N} [S(n) = S(m) \Rightarrow n = m]$
- A5. $\Phi(0) \wedge \forall_{n \in N} [\Phi(n) \Rightarrow \Phi(S(n))] \Rightarrow \forall_{n \in N} \Phi(n)$
- A6. $\forall_{n \in N} [n + 0 = n]$
- A7. $\forall_{n \in N} \forall_{m \in N} [n + S(m) = S(m + n)]$
- A8. $\forall_{n \in N} [n \times 0 = 0]$
- A9. $\forall_{n \in N} \forall_{m \in N} [n \times S(m) = (m \times n) + n]$

are the well-known first-order Peano arithmetic axioms (see, e.g., [1] or [2]). They establish the elementary theory of natural numbers and include three constants, i.e. 0 of an individual type, $N$ of a set type, and $S$ of a function type. $m$ and $n$ are variables of an individual type. $N$ is then the set of all natural numbers. $S(n)$ reads "successor of $n$", i.e. $S(n) = n + 1$ if an addition sign is used. A1 says that zero is a natural number, A2 says that every natural number has its successor being a natural number, and A3 says that zero is not the successor of any natural number, i.e. that zero is the smallest natural number. According to A4, every two natural numbers are equal if their successors are equal. A5 is an induction schema, which represents an infinite set of axioms with arbitrary formulae $\Phi$. A6 – A9 are recursive definitions of addition (A6 and A7) and multiplication (A8 and A9). According to the famous theorem of Gödel [3], the theory founded on A1 – A9 is *essentially incomplete.*

The scope of our considerations will be a spacetime which may have various curvatures (including flatness) in various regions and arbitrary numbers of spacelike and timelike dimensions. Our spacetime may be the whole universe or only its part. In this spacetime we choose a line $\mathcal{L}$ being either spacelike or timelike everywhere and such a coordinate system that $\mathcal{L}$ is one of its axes, say an axis of coordinates $x$. (Of course $\mathcal{L}$ may but need not be straight.) This means that every point of $\mathcal{L}$ is $(x,0,…,0)$ and all the coordinates $x$ are either spacelike or timelike, according to the kind of $\mathcal{L}$. Thus, the origin of our coordinate system is (0,0,…,0). We shall only be interested in the part of $L$ in which $x$'s are non-negative ($x \geq 0$). Out of these coordinates $x$, we choose an infinite sequence of the coordinates which we denote by $x_n$ and define as follows: $x_0 = 0$ (i.e. $(x_0,0,…,0)$ is the origin of our coordinate system); $x_n$ are numbered consecutively by means of all the natural numbers (i.e. there is a one-to-one mapping between $N$ and the set of all the coordinates $x_n$, and, in consequence, the set of all the points $(x_n,0,…,0)$); and $x_n < x_{n+1}$ for every $n \in N$. If $L$ is at least semi-infinite (semi- since $x_n \geq 0$), which then means that the universe is at least semi-infinite in at least one dimension, then there is no problem with constructing our sequence of $x_n$'s. If the universe is finite but there exists at least one continuous (or even only dense) dimension, then we choose $L$ along this dimension, and then we choose a segment of $L$ such that $0 \leq x < \alpha$. In this segment, we can define infinite sequences of $x_n$'s in many ways. For instance, we may, among other things, assume that

$$x_n := \alpha(1 - B^{-n}), \qquad B > 1, \qquad (1)$$

or

$$x_n := (2\alpha/\pi)\arctan Cn, \qquad C > 0, \qquad (2)$$

where $B$ and $C$ are real dimensionless constants. By choosing various values of $B$ ($> 1$) or $C$ ($> 0$) we can determine various positions of these $x_n$ in the segment $[0,\alpha)$.

Consider in our spacetime a physical quantity $Q$ values $q$ of which are determined at spacetime points (events). $Q$ may be a dynamical or kinematical quantity, it may be a scalar or a component of a tensor or anything else. On our line $L$ the values in question are of course $q(x,0,…,0)$ and there is a one-to-one mapping between $N$ and the set of all the values $q(x_n,0,…,0)$. We choose $Q$ such that

$$q(0,0,…,0) = 0, \qquad (3)$$

$$q(x_{n+1},0,…,0) = q(x_n,0,…,0) + D, \qquad D > 0, \qquad (4)$$

for every $n \in N$, where $D$ is a real constant. In virtue of (4) our $Q$ tends to infinity. Obviously, we can indicate or define many physical quantities satisfying (3) and (4) for various sequences of $x_n$'s under consideration. Note that physical quantities are frequently functions of other physical quantities. Among other things, some physical quantities tending to infinity are functions of physical quantities limited everywhere. Let us present two examples: (i) The Newtonian (or Coulombian) potential without its source is limited everywhere, whereas its reciprocal is a linear function tending to infinity. (ii) If we choose the spacetime distance from the origin of our coordinate system as a physical quantity (kinematical), then its values along $L$ are simply the coordinates $x$, and this quantity is limited in the segment $[0,\alpha)$. Assuming $Q$ such that $q(x,0,\ldots,0) = \tan(\pi x/2\alpha)$, we obtain $q(x_n,0,\ldots,0) = Cn$ in the case (2), i.e. a linear function of $n$ satisfying (3) and (4) with $D = C$.

Let us define
$$p_n := D^{-1}q(x_n,0,\ldots,0). \qquad (5)$$
$p_n$'s are therefore values of a dimensionless physical quantity correlated with our $Q$. In virtue of (5) and the one-to-one mappings mentioned above, there is a one-to-one mapping between $N$ and the set of all the values $p_n$, and from (3), (4), and (5) we obtain that $p_0 = 0$, $p_{n+1} = p_n + 1$ (i.e. $p_{n+1} = S(p_n)$), $p_0 \neq p_{n+1}$, $p_n + 0 = p_n$, $p_n + p_{m+1} = p_n + p_m + 1$, $p_n \times 0 = 0$, and $p_n \times p_{m+1} = p_n \times (p_m + 1) = (p_n \times p_m) + p_n$ for every $n \in N$ and every $m \in N$. In consequence, in A1 – A9 we can replace $m$ and $n$ with $p_m$ and $p_n$, respectively. In A1 – A9 modified in this way, $N$ acquires the meaning of a set of all the values $p_n$. As a result of this modification, we have obtained an *essentially incomplete system of sentences with a physical content.*

The above considerations can be summarized as follows. If we assume that the universe is infinite (or semi-infinite) even in only one dimension, or it is finite but includes a continuous (or even only dense) spacetime segment, then we must encounter independent sentences with a physical content. Any additional a priori assumption (non-contradictory of course) or empirical verification (or falsification) will not change this situation.

A situation different from those considered above occurs when one works on the assumption that a finite spacetime consists of the smallest segments of space and time. (Such a spacetime structure is commonly called granular or grainy.) Does therefore the assumption that the universe is finite and everywhere granular make it possible to produce a physical "theory of everything"? Perhaps it does, but a chance of the success seems to be slim.

**Acknowledgement.** I wish to thank Professor Marcin Mostowski for very helpful comments and discussions.